%% file: main.tex
\documentclass[journal,comcos]{IEEEtran}

\usepackage{ifpdf}
\ifpdf
\usepackage[T1]{fontenc}

\usepackage{amsmath,amsfonts,amssymb,mathrsfs}
\usepackage{graphicx}

\usepackage{epstopdf}
\DeclareGraphicsExtensions{.pdf,.eps,.png,.jpg,.mps}

\usepackage{algorithm}

\usepackage{algorithmicx}

\usepackage{algpseudocode}

\usepackage{bbm}

\usepackage{bm}

\usepackage[tight,footnotesize]{subfigure}

\usepackage{amssymb}

\usepackage{epsfig}

\usepackage{stfloats}

\usepackage{endnotes}

\usepackage{multirow}

\usepackage{adjustbox}

\usepackage{subfigure}

\usepackage{enumerate}

\usepackage{enumitem}

\DeclareFontFamily{OT1}{pzc}{}
\DeclareFontShape{OT1}{pzc}{m}{it}{<-> s * [1.10] pzcmi7t}{}
\DeclareMathAlphabet{\mathpzc}{OT1}{pzc}{m}{it}

\makeatletter
\def\user@resume{resume}
\def\user@intermezzo{intermezzo}
\newcounter{previousequation}
\newcounter{lastsubequation}
\newcounter{savedparentequation}
\makeatother

\begin{document}


\title{
Channel Tracking for Wireless Energy Transfer: A Deep Recurrent Neural Network Approach
}

\author{
Jae-Mo~Kang, Chang-Jae Chun, Il-Min~Kim,~\IEEEmembership{Senior~Member,~IEEE}, and Dong~In~Kim,~\IEEEmembership{Senior~Member,~IEEE} \vspace{-2em}
\thanks{J.-M. Kang, C.-J. Chun, and I.-M. Kim are with the Department of Electrical and Computer Engineering, Queen's University, Kingston, ON K7L 3N6, Canada (e-mail: jaemo.kang@queensu.ca; changjae.chun@queensu.ca; ilmin.kim@queensu.ca).}
\thanks{D. I. Kim is with the School of Information and Communication Engineering, Sungkyunkwan University (SKKU), Suwon, 16419, South Korea (e-mail: dikim@skku.ac.kr).}
}

\maketitle

\begin{abstract}
In this paper, we study channel tracking for the wireless energy transfer (WET) system,
which is practically a very important, but challenging problem.
Regarding the time-varying channels as a sequence to be predicted, we exploit the recurrent neural network (RNN) technique for channel tracking.
Particularly, combining the deep long short-term memory (LSTM) RNN with the deep feedforward neural network, we develop a novel channel tracking scheme for the WET system, which estimates the channel state information (CSI) at the energy transmitter based on the previous CSI estimates, and the current and previous harvested energy feedback information from the energy receiver.
Numerical results demonstrate the superior performance and effectiveness of the proposed scheme.
\end{abstract}

\begin{IEEEkeywords}
Channel tracking, deep learning, LSTM, recurrent neural network, wireless energy transfer.
\end{IEEEkeywords}

\IEEEpeerreviewmaketitle

\input{Tex/introduction}

\input{Tex/system_model}

\input{Tex/proposed_scheme}

\input{Tex/numerical_result}

\input{Tex/conclusion}


\input{Tex/reference}


\ifCLASSOPTIONcaptionsoff
  \newpage
\fi



\end{document}

%% file: Tex/introduction.tex
\vspace{-1em}
\section{Introduction}

Radio frequency (RF) wireless energy transfer (WET) is a very promising technology for energy-constrained networks
and it has broad applicability to wireless sensor networks, Internet-of-Things (IoT), etc \cite{R_Zhang}--\cite{Choi}.
To overcome the short WET distance and low WET efficiency due to the propagation loss,
in the WET systems, it is common to transfer the RF energy using the multiple-antenna technique, e.g., the energy beamforming \cite{R_Zhang}, \cite{J_Xu14}.
To transfer the RF energy as much as possible via the energy beamforming, the channel state information (CSI) must be perfectly or accurately known at the energy transmitter (ET),
which, however, is practically very challenging.
This is because almost all of the existing CSI acquisition algorithms developed for the wireless information transfer system are very difficult to realize at the energy receiver (ER)
due to the strictly limited (essentially, no) capability of RF-to-baseband conversion and baseband processing of the practical energy harvesting circuits \cite{Zeng15}, \cite{J_Xu}.

Considering the above critical issue, in \cite{Zeng15}--\cite{Kang}, several effective CSI acquisition schemes were developed for the WET system
under the assumption of quasi-static block fading:
the channels remain constant over the duration of each transmission block (i.e., several consecutive symbol times)
and vary from one block to another block independently.
In the real-world scenario, however, the fading channel varies dependently, e.g., as in the realistic continuous-state Markov channel model \cite{Gallager}.
Also, in practice, the channel varies even within a block, e.g., due to the mobility and/or the time-varying nature of the wireless environment.
In the realistic scenario of the time-varying channels, the schemes of \cite{Zeng15}--\cite{Kang} are neither effective nor efficient,
because the channels are estimated independently without considering the temporal dependencies or correlations of the channels
and the channel estimation task involving the complicated optimization process must be carried out frequently.

To address the above issue, in \cite{Choi}, an effective and efficient time-varying channel tracking scheme was developed for the WET system based on the Kalman filter.
It was demonstrated that this scheme works well in the realistic time-varying scenario \cite{Choi_2}.
However, in the channel tracking scheme of \cite{Choi},
the time-varying channels were indirectly estimated: the Gram (or one-sample correlation) matrices of the channels (rather than the channels themselves) were first estimated, from which the channels were then obtained via the rank-one approximation. Thus, the channel tracking performance of the scheme of \cite{Choi} is not satisfactory due to the approximation error.
Moreover, the scheme of \cite{Choi} is effective when the dynamics of the time-varying channels is linear. 
However, when the channel dynamics is nonlinear (as in \cite{nl_eq}), the scheme might not be as effective as in the linear case. 
To the best of our knowledge, all of the above critical issues have not been addressed in the literature.
This motivated our work.

In this paper, we propose a new channel tracking scheme for the WET system based on the recurrent neural network (RNN) technique,
which estimate the channel at the ET based on the previous channel estimates, and the current and previous harvested energy feedback information from the ER.
To this end, we construct a deep RNN architecture by combining the long short-term memory (LSTM) with the feedforward neural network (FNN).
The key idea of our proposed scheme is as follows: we regard the time-varying channels as a sequence to be predicted and we estimate the channels sequentially via the constructed deep RNN.
To the best of our knowledge, our work is the first to use the RNN technique for channel tracking in the literature.
The significant benefits of the proposed scheme over the conventional scheme of \cite{Choi} are as follows:
the channel tracking performance is significantly improved by successfully and effectively learning the temporal dependencies of the time-varying channels,
and the proposed scheme can be used whether the channel dynamics is linear or not.

\textit{Notation:} The transpose, the conjugate, the Hermitian transpose, the absolute value, the Euclidean norm, the trace, the element-wise multiplication, and the gradient operator with respect to $x$ are denoted by $(\cdot)^{T}$, $(\cdot)^{*}$, $(\cdot)^{H}$, $|\cdot|$, $\| \cdot \|$, $\textrm{Tr} (\cdot)$, $\odot$, and $\nabla_{x}$, respectively.

%% file: Tex/system_model.tex
\section{System Model and Conventional Scheme}

\subsection{System Model}

We consider a WET system composed of an ET and an ER, where the ET has $M$ antennas and the ER has a single antenna.
Let $\mathbf{h} (n) = [ h_{1} (n), \cdots, h_{M} (n) ]^{T} \in \mathbb{C}^{M \times 1}$ denote the channel vector from the ET to the ER at the $n$th symbol time,
where $h_{m} (n)$ is the channel coefficient from the $m$th antenna of the ET to the ER.
We will refer to the $n$th symbol time simply as time $n$.
We assume the time-varying channel, i.e., the channel varies from one time to another.
Also, we assume that each time duration is normalized to unity.
Thus, at time $n$, the harvested energy at the ER is given by $Q (n) = \zeta \big|  \mathbf{x}^{T} (n) \mathbf{h} (n)  \big|^{2}$,
where $0 < \zeta \leq 1$ is the energy conversion efficiency and $\mathbf{x} (n) \in \mathbb{C}^{M \times 1}$ is the energy signal sent from the ET.
From the Cauchy-Schwarz inequality, it follows that $Q (n) \leq \zeta  \| \mathbf{h} (n) \|^{2} \| \mathbf{x} (n) \|^{2}$,
where the equality is achieved when $\mathbf{x} (n) = \kappa \mathbf{h} (n)$ for some constant $\kappa$, which is called the energy beamforming technique \cite{R_Zhang}--\cite{Choi}.
This means that the channel vector $\mathbf{h}(n)$ must be estimated or tracked at the ET as accurately as possible to transfer the energy as much as possible
via the energy beamforming.

\vspace{-1em}
\subsection{Channel Tracking Problem}

For the channel estimation purpose, at each time, the ER feeds back the amount of harvested energy to the ET.
The feedback link is assumed to be error-free as in \cite{J_Xu}--\cite{Choi}.
The harvested energy feedback signal $r(n)$ received by the ET is then given by
\begin{align}
\label{yt}
r(n) = \zeta  \big|  \mathbf{x}^{T} (n) \mathbf{h} (n)  \big|^{2}  +  \eta(n) , \quad n=1,2,\cdots,
\end{align}
where $\eta(n) $ is the noise-plus-error term, which accounts for
the impairments in the measurement and quantization processes
such as the antenna noise, the rectifier noise, the measurement error, the feedback quantization error, etc.

Let $\hat{\mathbf{h}}(n)$ denote the estimate of $\mathbf{h}(n)$.
In this paper, the goal of the channel tracking is as follows: at time $n$, the current channel vector $\mathbf{h} (n)$ is estimated by the ET based on the estimates $\{ \hat{\mathbf{h}}(t) \}_{t=1}^{n-1}$ of the previous channel vectors,
and the current and previous harvested energy feedback information $\{ r(t) \}_{t=1}^{n}$.
This problem is a nonlinear estimation problem because the observations $\{ r(n) \}$ of (\ref{yt}) are all nonlinear in the variables $\{ \mathbf{h} (n) \}$ to be estimated.
Therefore, for the WET system, the channel tracking is generally very challenging.

\vspace{-1em}
\subsection{Conventional Channel Tracking Based on Kalman Filter}

In \cite{Choi}, a channel tracking scheme was developed for the WET system based on the Kalman filter.
Specifically, the harvested energy feedback information $\{ r(n) \}$ of (\ref{yt}) can be rewritten as
\begin{align}
\label{r2}
r(n) & = \zeta \textrm{Tr} \big( \mathbf{X} (n)  \mathbf{H} (n)   \big)  +  \eta(n), \quad \forall n,
\end{align}
where $\mathbf{H} (n) = \mathbf{h} (n) \mathbf{h}^{H} (n)$ and $\mathbf{X} (n) = \mathbf{x}^{*} (n)  \mathbf{x}^{T} (n) $
are the (rank-one) Gram matrices of the channel vector and the energy signal, respectively.
Note that in (\ref{r2}), the observations $\{ r(n) \}$ are all linear in $\{ \mathbf{H} (n) \}$.
Also, $\mathbf{X} (n)$ is known at the ET because $\mathbf{x} (n)$ is known.
Based on these results and the assumed channel model (i.e., autoregressive (AR) channel model), in \cite{Choi},
the channel Gram matrices $\{ \mathbf{H} (n) \}$ (rather than the channel vectors $\{ \mathbf{h}(n) \}$) were tracked using the Kalman filter.
Let $\hat{\mathbf{H}} (n)$ denote the estimate of $\mathbf{H} (n)$ obtained by the scheme of \cite{Choi}.
Then, using the eigenvalue decomposition, the rank-one approximation of $\hat{\mathbf{H}} (n)$ is given by $\hat{\mathbf{H}} (n) \approx \lambda_{1} \mathbf{u}_{1} \mathbf{u}_{1}^{H}$,
where $\lambda_{1} $ denotes the largest eigenvalue of $\hat{\mathbf{H}} (n)$ and $\mathbf{u}_{1}$ is the eigenvector corresponding to $\lambda_{1} $.
From this, the estimate of the channel vector $\mathbf{h}(n)$ can be obtained as $\hat{\mathbf{h}} (n) = \sqrt{\lambda_{1}} \mathbf{u}_{1} $.

Although the conventional scheme of \cite{Choi} can be used to track the channel vectors, it has certain limitations.
First of all, there exists the error in the rank-one approximation because the estimated channel Gram matrix $\hat{\mathbf{H}} (n)$ is not always guaranteed to be rank-one.
This approximation error deteriorates the channel tracking performance.
Furthermore, the scheme of \cite{Choi} was developed for the case when the dynamics of the time-varying channel vectors is linear,
i.e., when the current channel vector is represented by a linear combination of the previous ones.
Thus, when the channel dynamics is nonlinear (as in \cite{nl_eq}),
it is not clear whether the scheme of \cite{Choi} works as effectively as in the linear case.
To circumvent the above limitations intelligently and effectively, in the next section, we exploit the deep RNN technique.


%% file: Tex/proposed_scheme.tex
\vspace{-1em}
\section{Proposed Channel Tracking Based on Deep Recurrent Neural Network}

In this section, we develop a novel channel tracking scheme for the WET system by constructing a deep RNN architecture.
First, the structure and operation of the proposed scheme are presented.
Then we present the training procedure of the proposed scheme.


\begin{figure*}[t]
\centering
{
    \includegraphics[width=0.8\textwidth]{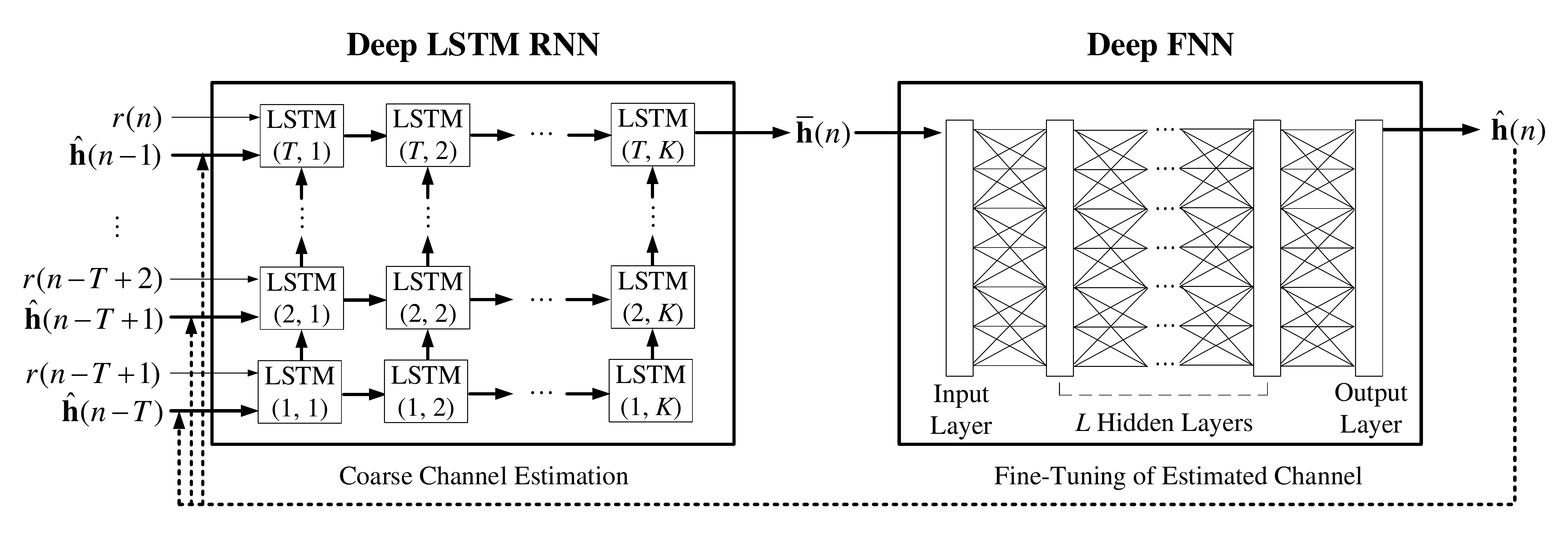}
    \caption{
    Structure of the proposed channel tracking scheme.
    }\vspace{-1em}
    \label{fig_1}
}
\end{figure*}

\vspace{-1em}
\subsection{Structure and Operation of the Proposed Scheme}

The structure of the proposed channel tracking scheme is shown in Fig. \ref{fig_1},
which is composed of two parts. In the first part, we construct a deep RNN by stacking the LSTMs
(which is referred to as the \textit{deep LSTM RNN})
to coarsely estimate the current channel vector.
In the second part, we construct a deep neural network by stacking the FNNs (which is referred to as the \textit{deep FNN}) to fine-tune the channel vector estimated by the deep LSTM network.
In the following, the network architecture and operation of each part is explained in detail.

\subsubsection{Coarse Channel Estimation with Deep LSTM RNN}
The LSTM is an advanced RNN, which is able to efficiently and effectively learn the long-term dependency or temporal property of the sequential data
by resolving the vanishing/exploding gradient problem involved in the the naive (or vanilla) RNN \cite{LeCun}, \cite{Goodfellow}.
In the LSTM, the following three gates are utilized: the forget gate $\mathbf{g}_{f} (n) \in \mathbb{R}^{q \times 1}$, the input gate $\mathbf{g}_{i} (n) \in \mathbb{R}^{q \times 1}$, and the output gates $\mathbf{g}_{o} (n) \in \mathbb{R}^{q \times 1}$,
which are given by
\begin{align}
\mathbf{g}_{j} (n) = \sigma \big(  \mathbf{W}_{j} \mathbf{i} (n) + \mathbf{U}_{j} \mathbf{c} (n-1) + \mathbf{b}_{j}  \big), \quad j \in \{ f, i, o \},
\end{align}
where $\mathbf{i} (n) \in \mathbb{R}^{d \times 1}$ and $\mathbf{c} (n) \in \mathbb{R}^{q \times 1}$ are the input and output of the node at time $n$, respectively,
and $\sigma(\cdot)$ is the activation function.
Also, $\mathbf{W}_{j} \in \mathbb{R}^{q \times d}$ and $\mathbf{U}_{j} \in \mathbb{R}^{q \times q}$ are the weight matrices
and $\mathbf{b}_{j} \in \mathbb{R}^{q \times 1} $ is the bias vector.
Note that the goal of the forget gate is to determine what redundant information is discarded; the goal of the input gate is to determine what new information is stored in the internal state; and the goal of the output gate is to determine what information to output.
Once the above three gates are computed, the LSTM updates its internal state $\mathbf{s} (n) \in \mathbb{R}^{q \times 1}$ as
\begin{align}
\mathbf{s} (n) & = \mathbf{g}_{f} (n) \odot \mathbf{s} (n-1) \nonumber\\
& \quad + \mathbf{g}_{i} (n) \odot \tanh \big(  \mathbf{W}_{s} \mathbf{i} (n) + \mathbf{U}_{s} \mathbf{c} (n-1) + \mathbf{b}_{s}  \big),
\end{align}
which determines how much the previous state information is kept and how much the current state information is allowed.
Then the output $\mathbf{c} (n) \in \mathbb{R}^{q \times 1}$ of the LSTM is given by
\begin{align}
\mathbf{c} (n) = \mathbf{g}_{o} (n) \odot \tanh \big(  \mathbf{s} (n)  \big).
\end{align}

In the proposed deep LSTM RNN, $T$ LSTMs are stacked vertically for the temporal processing and
$K$ LSTMs are stacked horizontally for the deep architecture, as shown in Fig. \ref{fig_1}. 
This enables to extract the high-level nonlinear temporal dependencies of the time-varying channels.
We take the input sequence of the first $T$ LSTMs (i.e., from the $(1, 1)$th to the $(T, 1)$th LSTMs) 
as the previous $T$ channel estimates $\{ \hat{\mathbf{h}}(n - t) \}_{t=1}^{T}$, and the current and previous $(T-1)$ harvested energy feedback information $\{ r(n - t) \}_{t=0}^{T-1}$.
Also, we use the output of the last LSTM (i.e., the $(T, K)$th LSTM), which is denoted by $\bar{\mathbf{h}} (n)$,
as the coarse (or initial) estimate of $\mathbf{h}(n)$.


\subsubsection{Fine-Tuning of the Estimated Channel with Deep FNN}
Note that the previous channel estimates $\{ \hat{\mathbf{h}}(n - t) \}_{t=1}^{T}$ (i.e., the inputs of the deep LSTM network) involve the estimation errors,
and thus, directly using the output $\bar{\mathbf{h}} (n)$ of the deep LSTM RNN as the channel estimate might not yield the satisfactory performance.
To mitigate the impacts of the estimation errors and to further improve the performance,
we propose to fine-tune the channel value estimated by the deep LSTM network using the deep FNN.

In the proposed deep FNN, there are the input layer, the $L$ hidden layers, and the output layer,
where the nodes in each layer are fully connected to those in the next layer.
The input of the deep FNN is the output of $\bar{\mathbf{h}} (n)$ of the deep LSTM network.
Also, the output is the estimate $\hat{\mathbf{h}}(n)$ of the current channel vector $\mathbf{h}(n)$ as the output,
which is fed into the deep LSTM RNN at the next time.
Thus, the number of the input nodes as well as the output nodes is given by $M$.
In the $k$th hidden layer, we employ $s_{k}$ nodes.
The output $\mathbf{o}_{k} \in \mathbb{R}^{s_{k} \times 1}$ at the $k$th hidden layer and the output $\mathbf{o}_{o} \in \mathbb{R}^{M \times 1}$ at each of the output nodes are given by
\begin{align}
\mathbf{o}_{j} = \varphi_{j}  \big(  \mathbf{V}_{j}  \boldsymbol{\iota}_{j}   + \boldsymbol{\beta}_{j}  \big),  \quad j \in \{ 1, \cdots, L, o  \},
\end{align}
where $\varphi_{j}$, $\boldsymbol{\iota}_{j}$, $ \mathbf{V}_{j}$, and $\boldsymbol{\beta}_{j}$ are the activation function, the input, the weight matrix, and the bias vector, respectively.

\vspace{-1em}
\subsection{Training of the Proposed Scheme}
\label{sec_training}






\begin{figure*}[t]
\begin{align}
\label{loss}
\mathcal{L}  \big( \{ \mathbf{h} (n) \}_{n=1}^{N} , \{ \hat{\mathbf{h}} (n) \}_{n=1}^{N} ; \theta \big)  = \frac{1}{N} \sum_{n=1}^{N} \big\| \mathbf{h} (n) - \hat{\mathbf{h}} (n)  \big\|^{2} = \frac{1}{N} \sum_{n=1}^{N} \big\| \mathbf{h} (n) - \mathcal{G}  \big( \mathcal{F} \big( \{ \hat{\mathbf{h}}(n - t) \}_{t=1}^{T},  \{ r(n - t) \}_{t=0}^{T-1}  ;  \theta_{\mathcal{F}}  \big)  ;  \theta_{\mathcal{G}}  \big)  \big\|^{2}.
\end{align}
\hrulefill\vspace{-1em}
\end{figure*}

In order to learn how to track the channels $\{ \mathbf{h} (n) \}$ as accurately as possible,
we use the mean square error (MSE) of the channel estimation given by (\ref{loss}) (at the top of the next page)
as the loss function for the training of our proposed scheme.
In (\ref{loss}), $N$ is the number of training samples and $\theta $ denotes the set of all the parameters (i.e., weights and biases) of the proposed scheme.
Also, $\mathcal{F} ( \cdot; \theta_{\mathcal{F}} )$ and $\mathcal{G} ( \cdot ;  \theta_{\mathcal{G}} )$ denote the overall mathematical functions of the proposed deep LSTM RNN and deep FNN, respectively,
where $\theta_{\mathcal{F}} $ and $\theta_{\mathcal{G}}$ are the sets of the parameters of the deep LSTM RNN and the deep FNN, respectively.
By minimizing the loss function $\mathcal{L} ( \cdot ; \theta )$ of (\ref{loss}) with respect to $\theta$, the proposed scheme can be trained and the parameters $\theta$ can be optimized.
To this end, we update the value of $\theta$ according to the gradient descent method as
\begin{align}
\label{gd}
\theta  \leftarrow \theta - \alpha \nabla_{\theta} \mathcal{L}  \big( \{ \mathbf{h} (n) \}_{n=1}^{N} , \{ \hat{\mathbf{h}} (n) \}_{n=1}^{N} ; \theta \big)
\end{align}
where $\alpha > 0$ is the learning rate or the step size of the update.

We note that the proposed scheme can be trained \textit{offline}.
Specifically, using the training samples of the channels $\{ \mathbf{h} (n) \}_{n=1}^{N}$ and the noises $\{ \eta(n) \}_{n=1}^{N}$ generated from the channel and noise processes, respectively,
the gradient step of (\ref{gd}) can be performed solely at the ET offline without the need for the ER to feed back the harvested energies online.
Thus, the proposed scheme is very useful and suitable for the practical applications because during the training, the ER does not need to carry out the feedback transmission,
which may require to consume a lot of energy at the ER and large amounts of time and bandwidth of the system.
Once the proposed scheme is trained, it can be used to efficiently and effectively track the channels at the ET in real-time based on the harvested energy feedback information.
Also, the proposed scheme can be used for the channel tracking whether the dynamics of the time-varying channels is linear or not,
which is in sharp contrast to the conventional scheme of \cite{Choi}.


%% file: Tex/numerical_result.tex
\vspace{-1em}
\section{Numerical Results}

In the numerical simulations, we consider a WET system with $M = 2$.
The channel vectors are generated according to the Gaussian-Markov model:
$\mathbf{h} (n) = \gamma \mathbf{h} (n - 1) + \mathbf{u} (n)$, $n=1,\cdots, 1000$,
where $\gamma \in [0,1]$ is the temporal fading correlation coefficient
and each element of $ \mathbf{u} (n) $ is the Gaussian process with zero mean and variance $(1 - \gamma^{2})$.
We set $\gamma = 0.998$. The elements of $\{\mathbf{x} (n)\}$ are drawn equally likely from $\{ -1, 1\}$
and those of $\{ \eta(n) \}$ are generated independently according to the Gaussian distribution with zero mean and variance $\varsigma^{2}$.
Thus, the signal-to-noise ratio (SNR) is given by $\frac{1}{\varsigma^{2}}$.
In our proposed scheme, we set $T = 2$, $K = 3$, and $L = 2$.
Also, the number of the hidden nodes in both the deep LSTM RNN and the deep FNN is set to be $20$.
The activation functions are set to $\sigma (x) = \tanh(x)$, $\varphi_{k} (x) = \max\{ x, 0 \}$, $k= 1,\cdots, K_{\rm FNN}$, and $\varphi_{o} (x) = x$.
The proposed scheme is trained using $N = 10^{4}$ training samples with the minibatch size of $10^{3}$ and the training epochs of $10^{2}$.
Then its performance is evaluated using $2 \times 10^{6}$ test samples that are different from the training samples.

\begin{figure}[t]
\centering
{
    \includegraphics[width=0.4\textwidth]{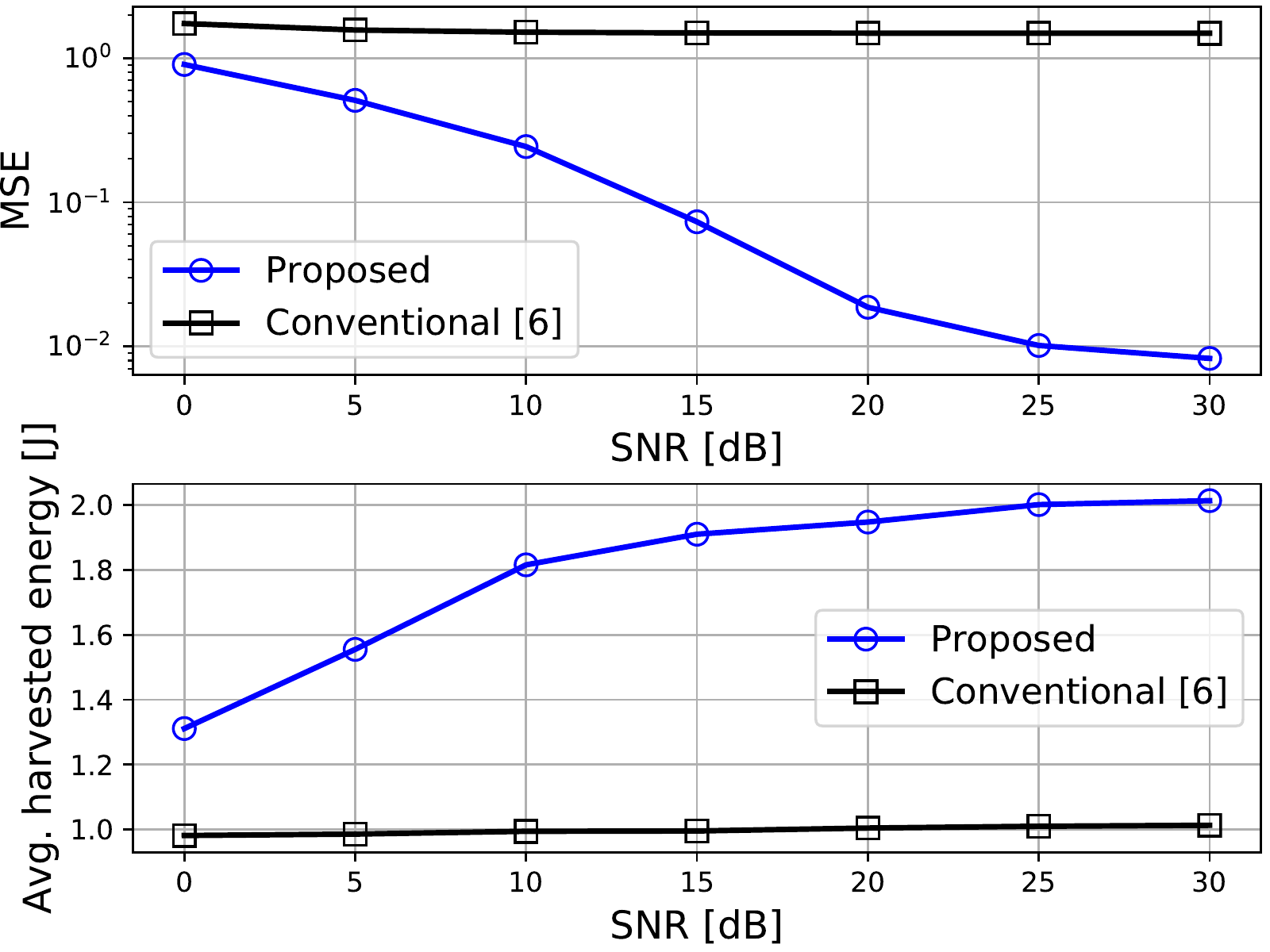}
    \caption{
    Channel estimation MSE and average harvested energy versus the SNR.
    }\vspace{-1em}
    \label{fig_1}
}
\end{figure}

\begin{figure}[t]
\centering
{
    \includegraphics[width=0.4\textwidth]{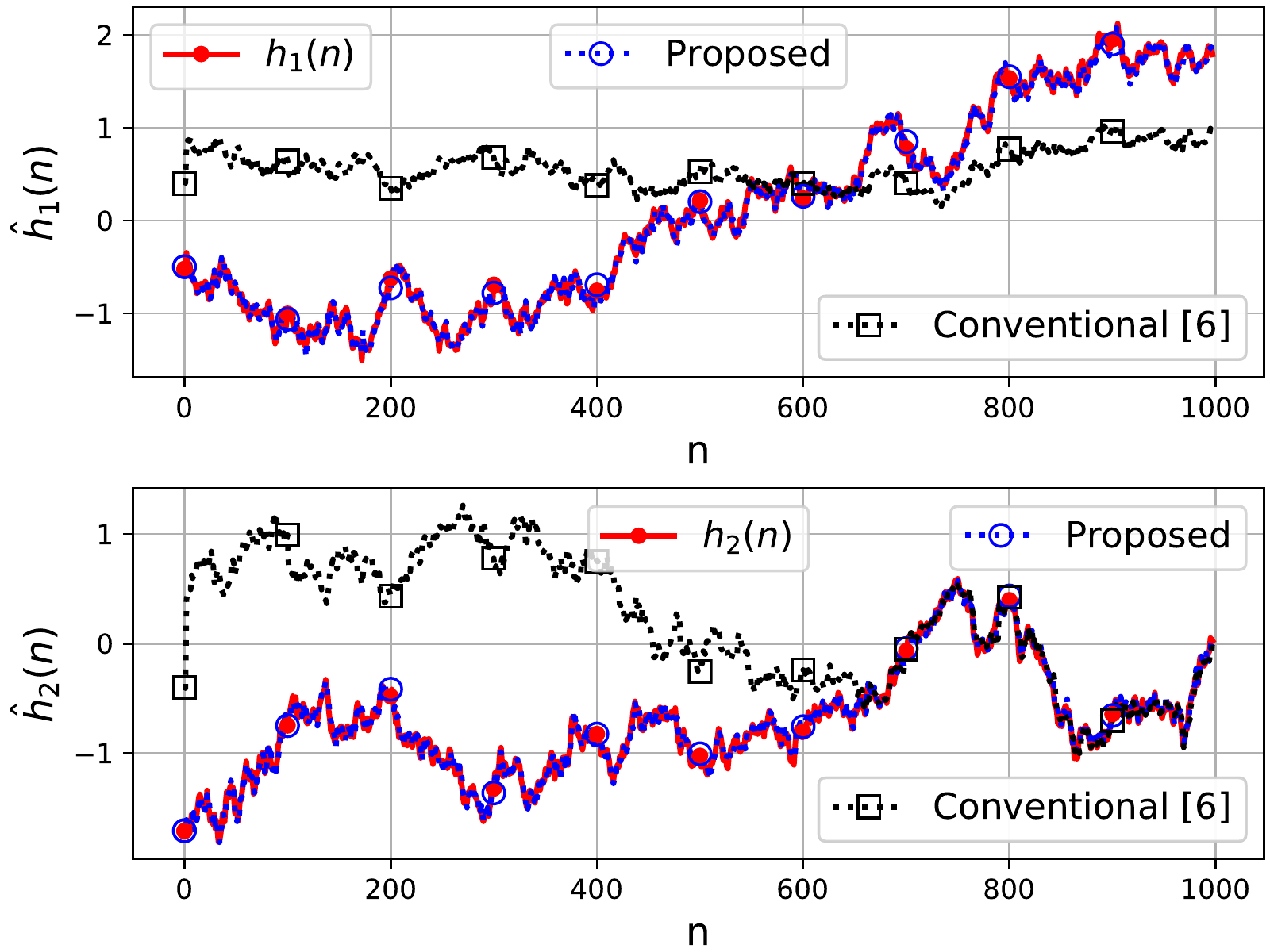}
    \caption{
    Channel estimates versus time $n$.
    }\vspace{-1em}
    \label{fig_2}
}
\end{figure}

\begin{figure}[t]
\centering
{
    \includegraphics[width=0.4\textwidth]{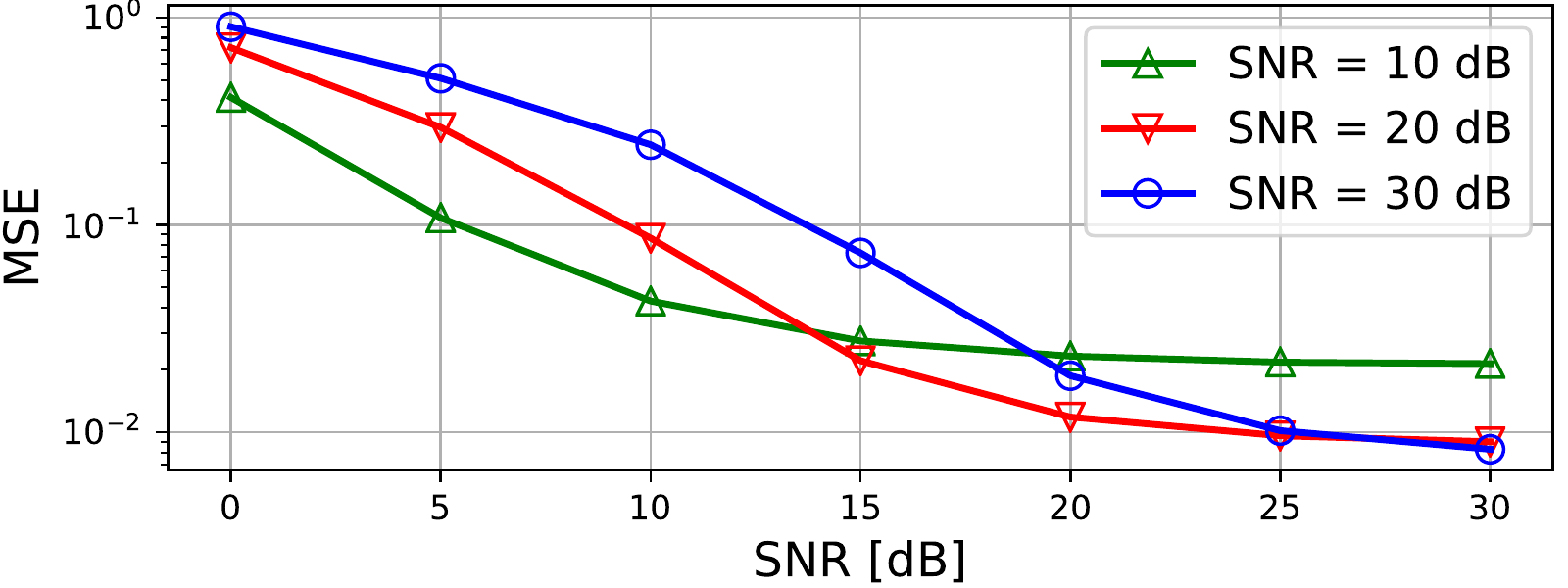}
    \caption{
    Performance of the proposed scheme with different SNR values for the training.
    }\vspace{-1em}
    \label{fig_3}
}
\end{figure}

In Fig. \ref{fig_1}, the channel estimation MSEs of the proposed and conventional schemes are shown versus the SNR, where the proposed scheme is trained at the SNR of 30 dB.
In this figure, the average harvested energies are also shown when the channel estimate is used as the energy beamforming, i.e., $\mathbf{x}(n) = \hat{\mathbf{h}} (n) / \| \hat{\mathbf{h}} (n) \|$, $\forall n$.
It can be seen that the proposed scheme has much lower MSE than the conventional scheme.
Thanks to this, the proposed scheme also harvests more energy than the conventional scheme.\footnote{
After the training, the computational complexity of the proposed scheme is $\mathcal{O} ( M Q T )$,
where $Q$ is the number of all the hidden and output nodes. On the other hand, the computational complexity of the conventional scheme is $\mathcal{O} ( M^{3} )$.
Thus, in the considered simulation setting, the complexity of the proposed scheme slightly higher than that of the conventional scheme.
}
In Fig. \ref{fig_2}, to see the channel tracking results, we plot the channel estimates obtained by the proposed and conventional schemes versus time $n$ when the SNR is 30 dB.
The results of Fig. \ref{fig_2} indicate that the proposed scheme is able to accurately track or predict the time-varying channels by learning the temporal property of the channels very well.
In Fig. \ref{fig_3}, we evaluate the performance of the proposed scheme when it is trained at different SNR values.
The MSEs at 10 dB, 20 dB, and 30 dB are lowest when the proposed scheme is trained at 10 dB, 20 dB, and 30dB, respectively.

%% file: Tex/conclusion.tex
\vspace{-1em}
\section{Conclusion}

A deep RNN based channel tracking scheme was proposed for the WET system.
The numerical results demonstrated that the proposed scheme considerably outperformed the existing scheme
in terms of the MSE and harvested energy.